# A journey into the tuneable antiferromagnetic spin textures of BiFeO$_3$

A. Haykal[1*], J. Fischer[2*], W. Akhtar[1], J.-Y. Chauleau[3], D. Sando[4], A. Finco[1], C. Carrétéro[2], N. Jaouen[5], M. Bibes[2], M. Viret[3], S. Fusil[2,6**], V. Jacques[1], V. Garcia[2]

[1]*Laboratoire Charles Coulomb, Université de Montpellier and CNRS, 34095 Montpellier, France*

[2]*Unité Mixte de Physique, CNRS, Thales, Univ. Paris-Sud, Université Paris-Saclay, 91767 Palaiseau, France*

[3]*SPEC, CEA, CNRS, Université Paris-Saclay, 91191 Gif-sur-Yvette, France*

[4]*School of Materials Science and Engineering, University of New South Wales, Sydney 2052, Australia*

[5]*Synchrotron SOLEIL, 91192 Gif-sur-Yvette, France*

[6]*Université d'Evry, Université Paris-Saclay, Evry, France*

*\*These authors contributed equally to this work*

*\*\*e-mail: stephane.fusil@cnrs-thales.fr*

**Antiferromagnetic thin films are currently generating considerable excitement for low dissipation magnonics[1] and spintronics[2,3,4]. However, while tuneable antiferromagnetic textures form the backbone of functional devices, they are virtually unknown at the submicron scale. Here we image a wide variety of antiferromagnetic spin textures in multiferroic BiFeO$_3$ thin films that can be tuned by strain and manipulated by electric fields through room temperature magnetoelectric coupling. Using piezoresponse force microscopy and scanning NV magnetometry in self-organized ferroelectric patterns of BiFeO$_3$, we reveal how strain stabilizes different types of non-collinear antiferromagnetic states (bulk-like[5,6] and exotic spin cycloids) as well as collinear antiferromagnetic textures. Beyond these local-scale observations, resonant elastic X-ray scattering confirms the existence of both types of spin cycloids. Finally, we show that electric-field control of the ferroelectric landscape induces transitions either between collinear and non-collinear states or between different cycloids, offering perspectives for the design of reconfigurable antiferromagnetic spin textures on demand.**

In ferromagnetic materials, spin textures are conventionally tweaked with a magnetic field. Antiferromagnetic spin textures, on the other hand, are intrinsically insensitive to external magnetic fields, calling for alternative control knobs to manipulate the antiferromagnetic order. The electrical manipulation of antiferromagnetism was recently demonstrated in non-centrosymmetric metallic antiferromagnets[3,7,8]; however, the spin orbit torque required to either switch by 90° or reverse by 180° the antiferromagnetic vector involves large current densities of the order of $10^6$-$10^7$ Acm$^{-2}$. Furthermore, the efficiency of this writing method faces limitations, since only a small fraction of antiferromagnetic domains is actually switched[9,10]. An optimal writing mechanism would demand low current densities (or ideally no current) to generate a complete reversal of antiferromagnetic domains or textures. Recent reports have for instance demonstrated that piezoelectric strain can provide low power control of antiferromagnetic memories[11,12].

In some materials possessing both antiferromagnetic and electrical orders, the magnetoelectric coupling is an additional resource expected to channel efficiently electric-field stimuli onto the antiferromagnetic order. Yet, fundamental ingredients deterministically governing the imprint of the ferroelectric order to the antiferromagnetic order remain poorly understood. Even in the archetypal room-temperature multiferroic[13], $BiFeO_3$, the details of the antiferromagnetic textures are virtually unknown at the scale of ferroelectric domains. To date, its complex antiferromagnetic order has been solely inferred from volume averaged techniques such as neutron diffraction, Mössbauer spectroscopy, or Raman spectroscopy. Depending on the strain, growth conditions and crystal orientation, the magnetic state of $BiFeO_3$ thin films can either show different types of non-collinear cycloids, canted G-type antiferromagnetic orders, or even a mixture of these[14,15]. More generally, examples of antiferromagnetic textures being imaged at the nanoscale are extremely scarce in the literature[16–18]. Here we bring deep insight into the strain-dependent interplay between the ferroelectric and antiferromagnetic orders at the local scale.

BiFeO$_3$ thin films were grown using pulsed laser deposition on various substrates (SrTiO$_3$, DyScO$_3$, TbScO$_3$, GdScO$_3$, SmScO$_3$) with a thin bottom electrode of SrRuO$_3$ (Methods). X-ray diffraction shows the high epitaxial quality of the films with Laue fringes (Fig. 1a-e) attesting for their coherent growth. All films display smooth surfaces with atomic steps, characteristic of a layer-by-layer growth (insets of Fig. 1a-e). The (001) BiFeO$_3$ peak evolves from the left to the right of the substrate (001) peak upon increase of the in-plane pseudo cubic lattice parameter of the substrate, as observed in the 2θ-ω scans. Reciprocal space maps indicate that the films are fully strained (Supplementary Figure 1) with only two elastic variants of the BiFeO$_3$ monoclinic phase (Fig. 1f-j). Their peak positions enable us to determine a strain value for each film ranging from -1.35% compressive strain to +0.50% tensile strain (Fig. 1k, Supplementary Figure 1 and Methods).

With this set of structurally equivalent BiFeO$_3$ thin films, distinguishable only by their strain level, we now focus on the evolution of the ferroelectric and magnetic textures (Figure 2). In BiFeO$_3$, the displacement of Bi ions relative to the FeO$_6$ octahedra gives rise to a strong ferroelectric polarisation along one of the <111> directions of the pseudo-cubic unit cell. The out-of-plane and in-plane variants of polarisation were identified in each sample using piezoresponse force microscopy (PFM; Methods). For all the samples, the as-grown out-of-plane polarisation is pointing downward, i.e. towards the bottom electrode (Supplementary Figure 2a). Figure 2a-e displays similar striped-domain structures with two in-plane ferroelectric variants corresponding to the elastic ones observed in reciprocal space maps[19]. In contrast to that of BiFeO$_3$ films grown on the scandates, the stripe-domain pattern on SrTiO$_3$ was defined by PFM (Supplementary Figure 3). All the samples can be considered as a periodic array of 71-degree domain walls, separated by two ferroelectric variants (Supplementary Figure 2). This ordered ferroelectric landscape greatly simplifies the exploration and interpretation of the magnetic configuration for each ferroelectric domain[20].

For each sample, the corresponding antiferromagnetic spin textures were imaged in real space with a scanning NV (nitrogen-vacancy) magnetometer[21] operated in dual-iso-B imaging mode (Fig. 2g-k, Methods). In the strain range of -1.35 to +0.05%, the NV images display a similar zig-zag pattern of periodic stray fields generated by cycloidal antiferromagnetic orders. More precisely, in each vertical ferroelectric domain (separated by dashed lines in Fig. 2g-j), we observe a single propagation direction of the spin cycloid. As the in-plane variant of polarisation rotates from one domain to another, the spin cycloid propagation direction rotates accordingly. This implies a one-to-one correspondence between the ferroelectric and antiferromagnetic domains. In contrast, for large tensile strain (+0.5%) corresponding to $BiFeO_3$ films grown on $SmScO_3$ substrates, the cycloidal order appears to be strongly destabilized (Fig. 2k). In this specific case, the ferroelectric periodicity is lost in the magnetic pattern, which may suggest a weaker magnetoelectric coupling as compared to other magnetic interactions. This strain dependence of the magnetic textures is reminiscent of previous works where antiferromagnetic order as a function of strain was studied by non-local techniques such as Mössbauer and Raman spectroscopies[14,15]. Indeed, a canted G-type antiferromagnetic order was identified for tensile strain over +0.5% and a cycloidal order from -1.6% to +0.5%.

In the present sample set, the magnetic image of $BiFeO_3$ films grown on $DyScO_3$ substrates (Fig. 2h) with -0.35% strain corresponds to the configuration already observed by Gross et al.[20]. The 90-degree in-plane rotation of the ferroelectric polarisation imprints the 90-degree in-plane rotation of the cycloidal propagation direction. This corresponds to one of the three bulk-like cycloids (cycloid I) with propagation vectors contained in the (111) plane orthogonal to the polarisation (Fig. 3a-b). Among them, the observed $k_1$ vector lies in the (001) plane of the film, for both ferroelectric variants (Fig. 2h). For lower compressive strain (-0.10%, $TbScO_3$), the magnetic configuration is found to be identical, also corresponding to the bulk-like cycloid (cycloid I, $k_1$).

A subtle change of the strain towards the tensile side (+0.05%, GdScO$_3$) influences the magnetic landscape. Indeed, the spin texture can no longer be explained by the bulk-like cycloid as the zig-zag angle is not at 90 degrees anymore, but 120 ± 5 degrees (Fig. 2j). There are previously reported indications of exotic spin cycloids for (001)BiFeO$_3$ films grown under small tensile strain (+0.2%) [14,15]. In these works, Mössbauer and nuclear resonant scattering data suggested a propagation vector contained in the ($\bar{1}$10) plane[14,15]. This result was recently supported by neutron diffraction experiments on Co-doped BiFeO$_3$ films grown on SrTiO$_3$(110), where the propagation direction of the spin cycloid was found to be along the [11$\bar{2}$] direction[22]. Guided by these observations, here we consider three possible propagation directions ($k'_1, k'_2, k'_3$) for the cycloid II; namely along [$\bar{2}$11], [1$\bar{2}$1], and [11$\bar{2}$], respectively (Fig. 3c-d). In the case of BiFeO$_3$ thin films on GdScO$_3$ substrates (Fig. 2j), the angle of the zigzag pattern is only compatible with alternating $k'_1$, $k'_2$ propagation vectors, giving rise to an angle of 127 degrees, as projected on the film surface. Surprisingly, a similar scenario takes place for large compressive strain (-1.35%, SrTiO$_3$) as the zigzag angle (Fig. 2g) is the same as for BiFeO$_3$ grown on GdScO$_3$. Such unprecedented real-space observations of this cycloid II in BiFeO$_3$ for both compressive *and* tensile strain calls for further theoretical input to explain the interplay between strain and antiferromagnetic textures.

Complementary macroscopic investigations were performed by X-ray resonant elastic scattering on BiFeO$_3$ samples[23,24] grown on both DyScO$_3$ (cycloid I) and GdScO$_3$ (cycloid II) substrates (Fig. 4a,c). As the spin cycloid is a periodic magnetic object, it gives rise to a diffracted pattern at the Fe resonant L-edge. In order to select the diffracted signal of magnetic origin, the difference between left and right circularly polarized light is plotted as a dichroic diffracted pattern (Fig. 4a, red and blue contrasts correspond to positive and negative dichroism, respectively). In both diagonals from the specular spot, the inverted contrast between +$q$ and -$q$ spots is a signature of chirality. Indeed, BiFeO$_3$ spin cycloids in which spins rotate in a plane defined by the polarisation (*P*) and the propagation vector ($k$) are chiral objects.

For BiFeO$_3$ thin films grown on DyScO$_3$, considering two orthogonal cycloid propagation directions (red arrows in Fig. 4a) with identical periods gives rise to two orthogonal lines of diffracted spots, defining a square diffracted pattern. The fine structure of this pattern is rendered more complex by additional spots coming from the modulation of the magnetic periodicity by the ferroelectric one[25]; however, here our focus is on the cycloid propagation direction and periodicity. The spacing between the +*q* and -*q* spots corresponds to a cycloid period of 72 ± 5 nm for both spin cycloids with $k_1$ propagation vector. Consistently at the local scale, the combination of PFM and scanning NV magnetometry allows to identify the relative orientation of the ferroelectric polarisation (*P*, grey arrows in Fig. 4b) and cycloid propagation direction ($k_1$, red arrows in Fig. 4b) on both sides of a domain wall. Thus, both microscopic real-space experiments and macroscopic reciprocal-space ones are pointing toward a single cycloidal vector ($k_1$) in BiFeO$_3$ thin films with moderate compressive strain.

In contrast, for BiFeO$_3$ films grown on GdScO$_3$ imposing slight tensile strain, the dichroic diffracted pattern is no longer square but rectangular (Fig. 4c). Hence, we preclude the above-mentioned scenario with two bulk-like (cycloid I) orthogonal vectors. The two diagonals of the rectangular pattern (green arrows in Fig. 4c) form an angle of about 110 ± 5 degrees, in accordance with the typical angles observed in NV magnetometry images. The only plausible scenario therefore corresponds to two types of ferroelectric domains respectively harbouring alternating $k_1'$ and $k_2'$ propagation vectors of the cycloid II, as observed in real space (Fig. 4d). These two cycloid propagation variants appear to be energetically degenerated and favoured over the more out-of-plane $k_3'$ vector (Fig. 3c). Consequently, for both tensile and compressive strain, these cycloidal BiFeO$_3$ films exhibit a one-to-one imprint from ferroelectric to antiferromagnetic orders.

Beyond the observations on pristine configurations of ferroelectric domains in which the cycloid propagation is locked onto the polarisation, we now manipulate the ferroelectric order, with an aim to design on demand antiferromagnetic landscapes with electric fields. We first use PFM to draw micron-size ferroelectric domains by virtue of the so-called trailing field[26–28]. NV magnetometry is then performed on these artificial domains to reveal the corresponding magnetic textures (Fig. 5). For strain states ranging from -0.35 to +0.50%, single ferroelectric domains always correspond to a spin cycloid with a single propagation vector. For $BiFeO_3$ films grown on $DyScO_3$ (-0.35%, Fig. 5a) or $TbScO_3$ (-0.10%, Fig. 5b), the spin cycloid propagates perpendicularly to the ferroelectric polarisation. This implies that the in-plane $k_1$ propagation is still favoured, switching from two pristine cycloid Is to a single written cycloid I. Interestingly, the spin cycloid period λ decreases from about 78 ± 5 nm in the pristine (two domain) state to 65 ± 2 nm for the switched (single domain) state. In single domains, the spin cycloid period thus appears closer to the one observed in bulk $BiFeO_3$ ($λ_{bulk}$ = 64 nm, Ref. [6]), suggesting that periodic electric/elastic boundary conditions influence the cycloid period.

For $BiFeO_3$ films grown on $GdScO_3$ (+0.05%, Fig. 5c), the spin cycloid propagates horizontally, i.e. at 45 degrees from the in-plane polarisation variant of the single ferroelectric domain. This implies that the cycloid I out-of-plane propagation vector ($k_2$, Fig. 3a-b) is selected, corresponding to a switching from two cycloid IIs ($k'_1, k'_2$) to a single cycloid I ($k_2$). In addition, the apparent cycloid period of 92 ± 3 nm in the single domain is compatible with its projection onto the sample surface ($λ_{surf}$ = √2 × λ), giving rise to an intrinsic period of λ = 65 ± 2 nm, close to the bulk value. These experiments on single domains suggest that strain primarily has an influence on the direction of the bulk-like cycloid propagation (in-plane for compressive and out-of-plane for tensile strains). In the case of $BiFeO_3$ films grown on $SmScO_3$ (+0.50%, Fig. 5d), the cycloid is observed to propagate in a direction almost parallel to the in-plane variant of polarisation. Considering the three vectors of each cycloid type (Fig. 3), this is only compatible with the $k'_3$ propagation vector of cycloid II. In this case, we find an apparent cycloid period of 146 ± 5 nm leading to an intrinsic period of 84 ± 3 nm ($\lambda_{surf} = \sqrt{3} \times \lambda$). The enhanced period compared to the bulk value is here attributed to the significant tensile strain of BiFeO$_3$ films grown on SmScO$_3$ (Ref. [15]). In this latter example, we have demonstrated electric-field switching from a G-type antiferromagnetic order to a cycloidal state.

In this work, we have imaged multiple antiferromagnetic landscapes in real-space such as bulk-like cycloids, exotic cycloids, and G-type collinear orders, depending on the epitaxial strain of BiFeO$_3$. The exotic cycloid is unexpectedly stabilized for both compressive and tensile strain. Combining multiple scanning probe techniques, we provide direct correspondence between ferroelectric domains and complex antiferromagnetic textures. These local observations are supported by macroscopic resonant X-ray scattering on both types of cycloids. While often omitted in the literature of BiFeO$_3$ films[29], we find that only the cycloidal state promotes a full imprint between both ferroic orders in the native striped-domains as well as in artificially-designed single domains. The electric field enables toggling either from one type of cycloid to another or from collinear to cycloidal states. More specifically, we are now able to electrically design single spin cycloids on demand with controlled propagation either in the plane or out of the film plane. This fully mastered magnetoelectric system is an ideal playground to investigate reconfigurable low-power antiferromagnetic spintronic or magnonic architectures at room temperature.

## Acknowledgments

We acknowledge support from the French Agence Nationale de la Recherche (ANR) through the PIAF project, the European Research Council (ERC-StG-2014, Imagine), the EU Quantum Flagship project ASTERIQS (820394) and the European Union's Horizon 2020 research and innovation programme under the Marie Sklodowska-Curie grant agreement No 846597. This work was supported by a public grant overseen by the ANR as part of the 'Investissement d'Avenir' programme (LABEX NanoSaclay, ref. ANR-10-LABX-0035). We also acknowledge the company QNAMI for providing all-diamond scanning tips containing single NV defects.

## Authors contributions

V.G., S.F. and V.J conceived and coordinated the experiment. J.F. and C.C. prepared the samples. J.F. carried out the X-ray diffraction experiments and analysed the structural properties of the samples with D.S. and V.G. J.F., S.F. and V.G. performed the piezoresponse force microscopy experiments. A.H., W.A., A.F. and V.J. conducted the scanning NV magnetometry experiments. J.-Y.C., N.J. and M.V. performed the resonant X-ray scattering experiments. V.G. and S.F. wrote the manuscript with inputs from J.F., M.B., D.S. and V.J. All the authors discussed the data and commented the manuscript.

**Figure legends**

**Figure 1. Strain-engineered epitaxial BiFeO₃ thin films. a-e**, 2θ-ω X-ray diffraction scans of BiFeO₃ (BFO) films grown on SrTiO₃ (STO) (**a**), DyScO₃ (DSO) (**b**), TbScO₃ (TSO) (**c**), GdScO₃ (GSO) (**d**) and SmScO₃ (SSO) (**e**) substrates. The insets are 3 × 3 μm² topography images acquired by atomic-force microscopy on the same films, showing atomic steps and terraces. The z-scale is 4 nm. **f-j**, Corresponding reciprocal space maps along the different (113) substrate peaks, showing in each case two elastic domains for BiFeO₃, i.e. (203) and (023). **k**, Sketch of the evolution of the calculated epitaxial strain in BiFeO₃ as a function of the substrate. The scandate and BiFeO₃ crystallographic peaks are defined in a monoclinic cell.

**Figure 2. Strain vs. magnetic textures on striped ferroelectric domains. a-e**, In-plane PFM phase images of BiFeO₃ films grown on SrTiO₃ (**a**), DyScO₃ (**b**), TbScO₃ (**c**), GdScO₃ (**d**) and SmScO₃ (**e**) substrates. **f**, Sketch of the evolution of the epitaxial strain in BiFeO₃ as a function of the substrate. **g-k**, NV magnetometry images corresponding to the ferroelectric domains depicted in (**a-e**).

**Figure 3. Sketches of the different types of spin cycloids in BiFeO₃. a,b**, Bulk-like spin cycloid (cycloid I) with the 3 possible propagation vectors for each polarisation variant in 3D view (**a**) and top view (**b**). **c,d**, The exotic spin cycloid (cycloid II) with propagation vectors along the three <11$\bar{2}$> directions in 3D view (**c**) and top view (**d**).

**Figure 4. The two types of spin cycloids in real and reciprocal spaces. a**, Resonant X-ray elastic scattering at the Fe L-edge for BiFeO₃ grown on DyScO₃. The square pattern indicates a bulk-like cycloid (cycloid I) with propagation vectors aligned 90 degrees from each other. **b**, Corresponding NV magnetometry image zoomed in, with the propagation vectors sketched for both polarisation variants. **c**, Resonant X-ray elastic scattering at the Fe L-edge for BiFeO₃ grown on GdScO₃. The rectangular pattern corresponds to the cycloid II with propagation vectors lying at 110 ± 5 degrees from each other. **d**, Corresponding NV magnetometry image zoomed in, with the propagation vectors sketched for both polarisation variants.

**Figure 5. Magnetic textures in single ferroelectric domains as a function of strain. a**-**d**, NV magnetometry images in single ferroelectric domains defined preliminarily by PFM for BiFeO$_3$ thin films grown on DyScO$_3$ (**a**), TbScO$_3$ (**b**), GdScO$_3$ (**c**), and SmScO$_3$ (**d**). The corresponding strain are depicted in the first row and the second row indicates the evolution of the magnetic textures from striped domains to single ferroelectric domains. The propagation vector of the spin cycloid relative to the ferroelectric polarisation is sketched below each image.

**Supplementary Figure 1. Reciprocal space maps on (103), (013), ($\bar{1}\bar{1}3$) substrate peaks.** BFO, STO, DSO, TSO, GSO and SSO stand for BiFeO$_3$, SrTiO$_3$, DyScO$_3$, TbScO$_3$, GdScO$_3$ and SmScO$_3$, respectively. The two colours stand for the two elastic domains of the BiFeO$_3$ thin films.

**Supplementary Figure 2. Striped ferroelectric domains in BiFeO$_3$ with 71-degree domain walls. a**, Out-of-plane PFM phase image of a BiFeO$_3$ film grown on TbScO$_3$(110). The homogeneous bright signal indicates a downward polarisation. **b**, Corresponding in-plane PFM phase image. The striped-domain structure corresponds to two polarisation variants (grey arrows). **c**, Sketch of the 71-degree domain wall structure.

**Supplementary Figure 3. Artificial stripes designed by PFM on BiFeO$_3$ thin films grown on SrTiO$_3$. a**, Out-of-plane PFM phase change from domains pointing downwards (bright contrast) to domains pointing upwards (dark contrast). **b**,**c**, This writing scheme is accompanied by a change in the arrangement of the in-plane polarisation variants from the native mosaic-like pattern (**b**) to a stripe-domain pattern (**c**).

## Methods

**Sample fabrication.** BiFeO$_3$ thin films were grown by pulsed laser deposition on various substrates using a KrF excimer laser (248 nm) with a fluence of 1 J/cm$^2$. Prior to film growth, the scandate substrates (DyScO$_3$, TbScO$_3$, GdScO$_3$, SmScO$_3$) were ex situ annealed for 3 hours at 1000°C under flowing oxygen. The SrTiO$_3$ substrate was chemically etched with a buffered HF diluted solution before following the same annealing procedure. For all the samples, a SrRuO$_3$ bottom electrode (3-5 nm) was first grown at 660°C under 0.2 mbar of oxygen pressure with a laser repetition rate of 5 Hz. The BiFeO$_3$ thin film (30-60 nm) was subsequently grown at the same temperature under 0.36 mbar of oxygen pressure and a repetition rate of 2 Hz. Following the growth of the bilayer, the samples were cooled down to room temperature under an oxygen pressure of 300 mbar.

**Structural characterisations.** The structural properties of the films were determined by X-ray diffraction (XRD) using a Panalytical Empyrean diffractometer equipped with a hybrid monochromator for Cu K$_{\alpha 1}$ radiation and a PIXcel3D detector. Full 2θ-ω XRD scans (not shown) indicate that all films are single phase with a monoclinic (001) orientation. To get more insights into the elastic domains and strain of the films, we carried out reciprocal space maps (RSMs) around the (103), (013), (113), and ($\bar{1}\bar{1}3$) substrate peaks (Fig. 1g-k and Supplementary Figure 1). The (110) orthorhombic scandates (XSO with X = Dy, Tb, Gd, Sm) are all described in a (001) monoclinic notation for simplicity[30]. The RSMs are consistent with only two monoclinic ferroelastic variants of BiFeO$_3$ with the following epitaxial relationship: (001)BFO||(001)XSO, [100]BFO||[110]XSO (green) and (001)BFO||(001)XSO, [100]BFO||[1$\bar{1}$0]XSO (blue). The same epitaxial relationship is established for BiFeO$_3$ films grown on cubic (001)SrTiO$_3$ substrates. The BiFeO$_3$ thin films are fully strained by the substrates as indicated by the alignment of the in-plane reciprocal peaks with the (103) and (013) substrate peaks (Supplementary Figure 1). The monoclinic cell parameters ($a_m, b_m, c_m, \beta$) of each BiFeO$_3$ film were calculated independently from the peak positions around the (113) and ($\bar{1}\bar{1}3$) RSMs of XSO. The strain values were then estimated by comparing the average in-plane lattice parameter with the volume of the unit-cell as:

$$\varepsilon = \frac{\sqrt{\frac{a_m \times b_m}{2}} - \sqrt[3]{\frac{V}{2}}}{\sqrt[3]{\frac{V}{2}}}, \text{ where } V = a_m \times b_m \times c_m \times \beta$$

Considering the small deviation from the cubic unit cell, the description of the ferroelectric and magnetic properties is done in the pseudo-cubic perovskite lattice for the sake of simplicity.

**Piezoresponse force microscopy.** The experiments were conducted with an atomic force microscope (Nanoscope V multimode, Bruker) and two external lock-in detectors (SR830, Stanford Research) for the simultaneous acquisition of in-plane and out-of-plane responses. An external ac source (DS360, Stanford Research) was used to excite the $SrRuO_3$ bottom electrode at a frequency of 35 kHz while the conducting Pt-coated tip was grounded. We used stiff cantilevers (40 N/m) for accurate out-of-plane detection and softer ones (3-7 N/m) for the in-plane detection. In all the $BiFeO_3$ samples, the as-grown out-of-plane signal is homogeneous (Supplementary Figure 2) indicating a uniform out-of-plane component of polarisation pointing downwards, i.e. towards the $SrRuO_3$ bottom electrode.

**Scanning NV magnetometry.** Scanning-NV magnetometry was performed under ambient conditions with commercial all-diamond scanning-probe tips containing single NV defects (QNAMI, Quantilever MX). The tip was integrated into a tuning-fork-based atomic force microscope (AFM) combined with a confocal microscope optimized for single NV defect spectroscopy. Magnetic fields emanating from the sample are detected by recording the Zeeman shift of the NV defect's electronic spin sublevels through optical detection of the electron spin resonance[21].

The scanning-NV magnetometer was operated in the dual-iso-B imaging mode by monitoring the signal S=PL($v_2$)–PL($v_1$), corresponding to the difference of photoluminescence (PL) intensity for two fixed microwave frequencies, $v_1$ and $v_2$, applied consecutively at each point of the scan through a gold stripline antenna directly fabricated onto the BFO sample by e-beam lithography[21]. Experiments were performed with a NV-to-sample distance of 60 nm and a bias magnetic field of 2 mT applied along the NV quantization axis. The standard error of the cycloid period measurement is limited by the calibration of the scanner.

**Resonant X-ray elastic scattering.** Resonant X-ray scattering measurements were performed at the Fe L and O K edges using the RESOXS diffractometer[31] at the SEXTANTS beamline[32] of the SOLEIL synchrotron. Data were collected using nearly fully circular left (CL) and right (CR) X-ray polarisations delivered by the HU44 Apple2 undulator located at the I14-M straight section of the storage ring.

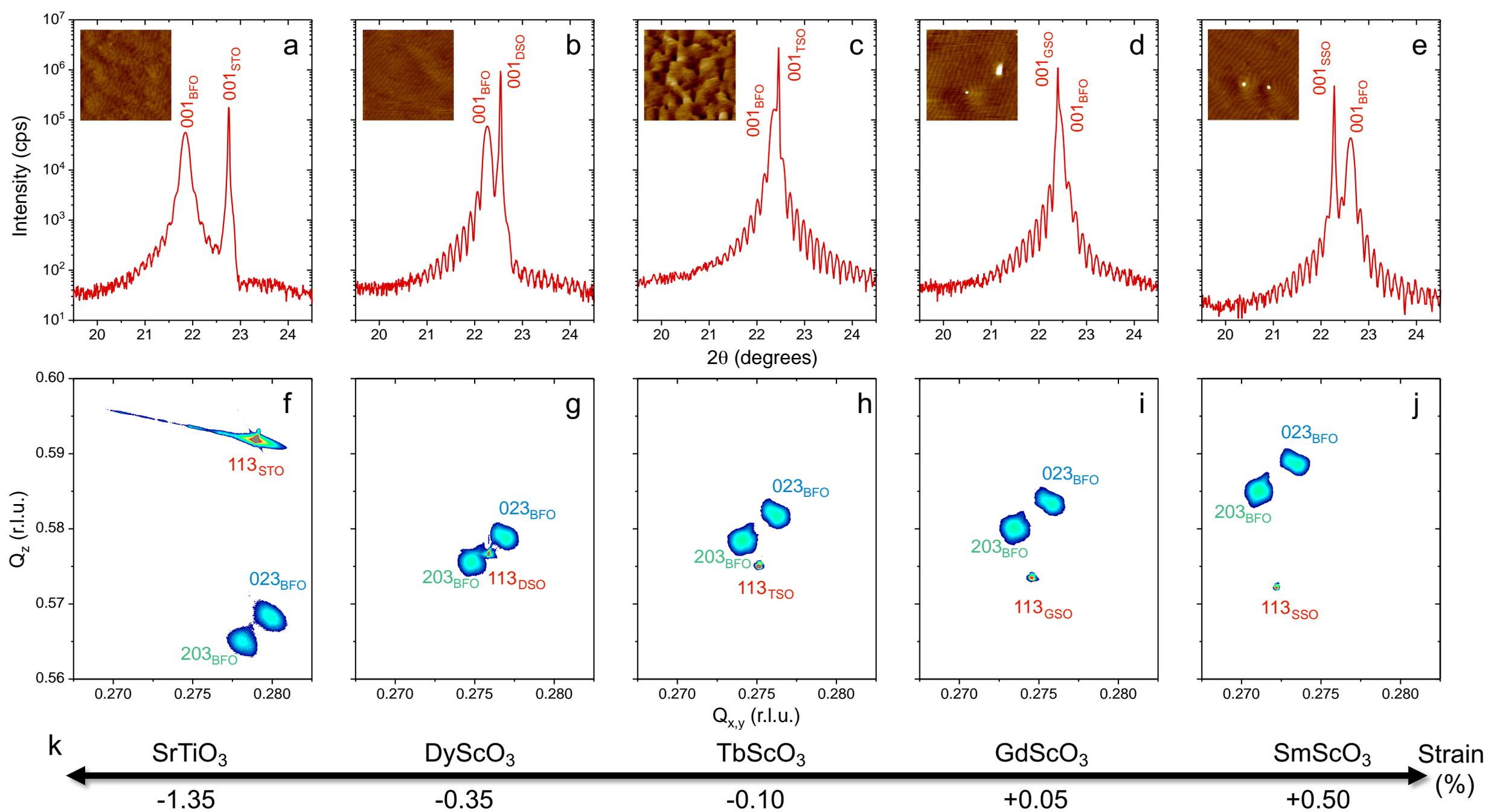

Figure 1: Strain-engineered epitaxial BiFeO₃ thin films

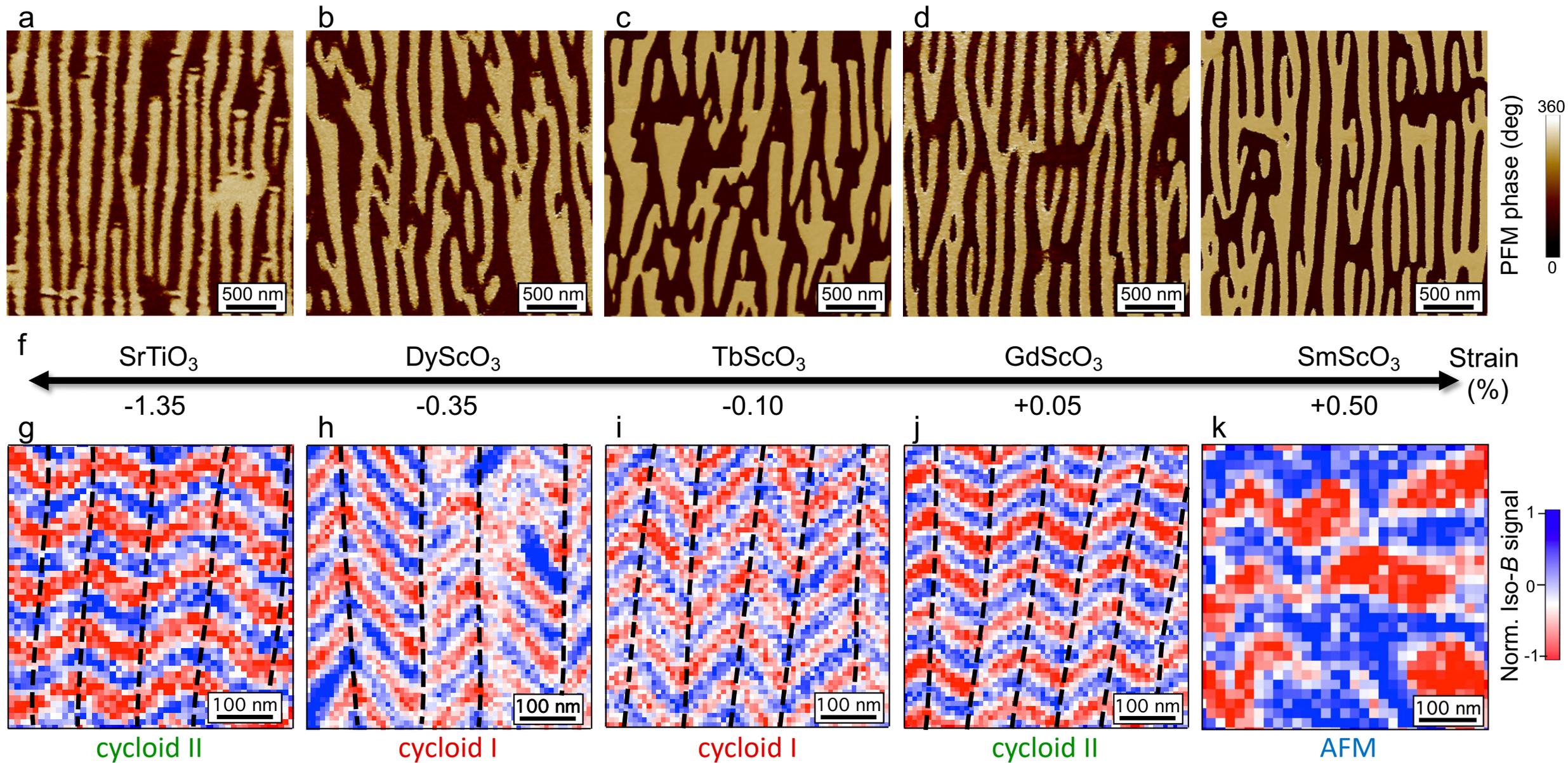

Figure 2: Strain vs. magnetic textures on striped ferroelectric domains

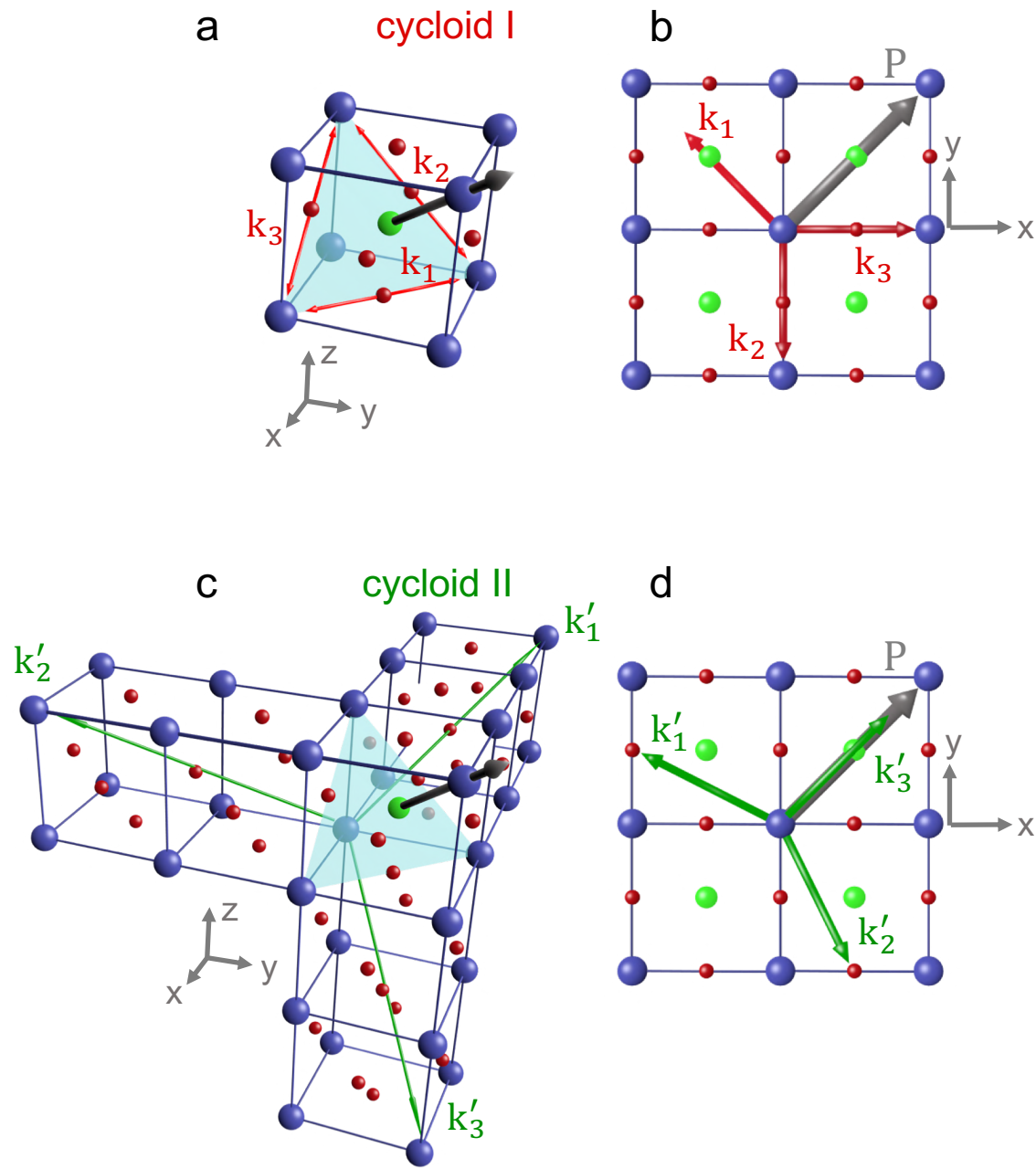

Figure 3: Sketches of the different types of spin cycloids in BiFeO$_3$

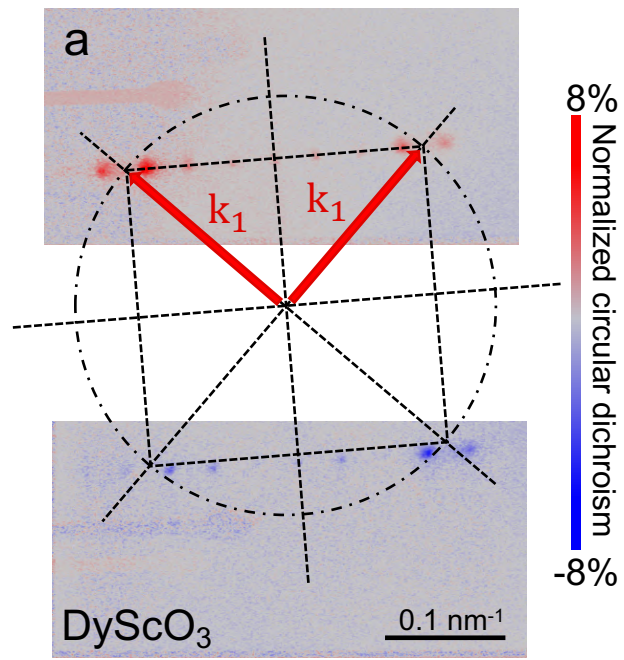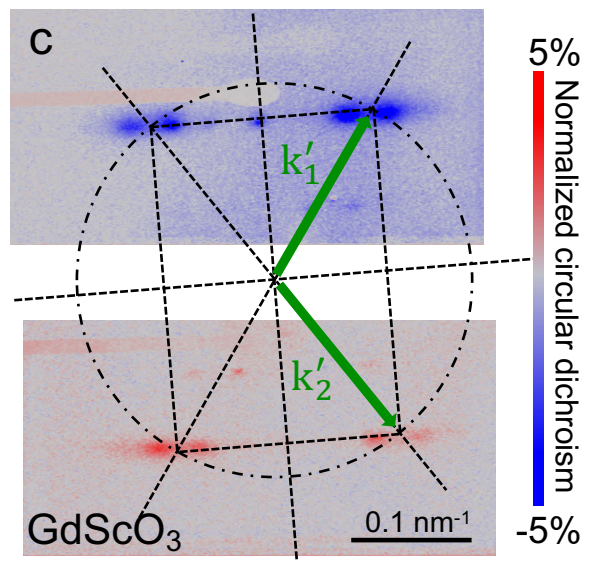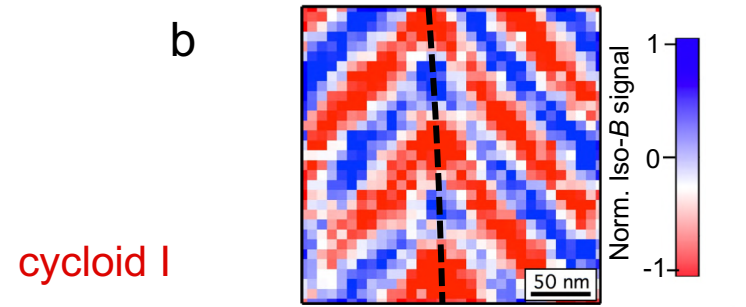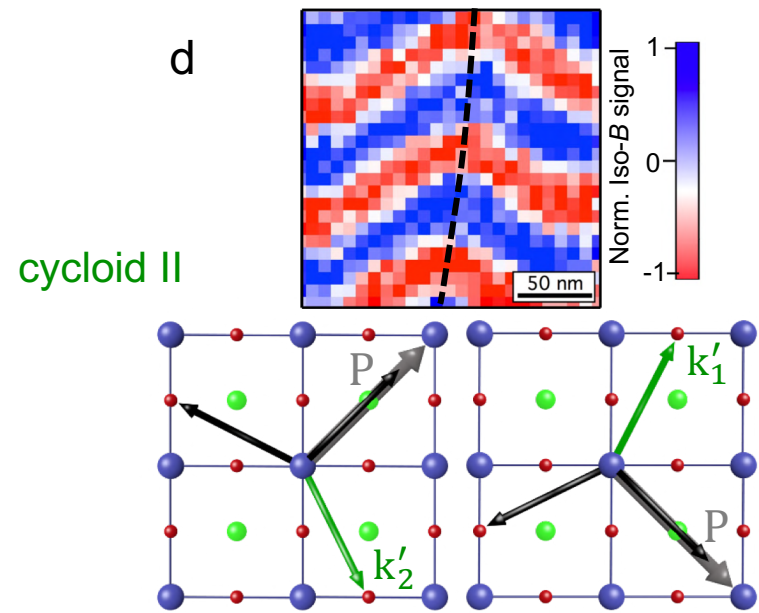

Figure 4: The two types of spin cycloids in real and reciprocal spaces

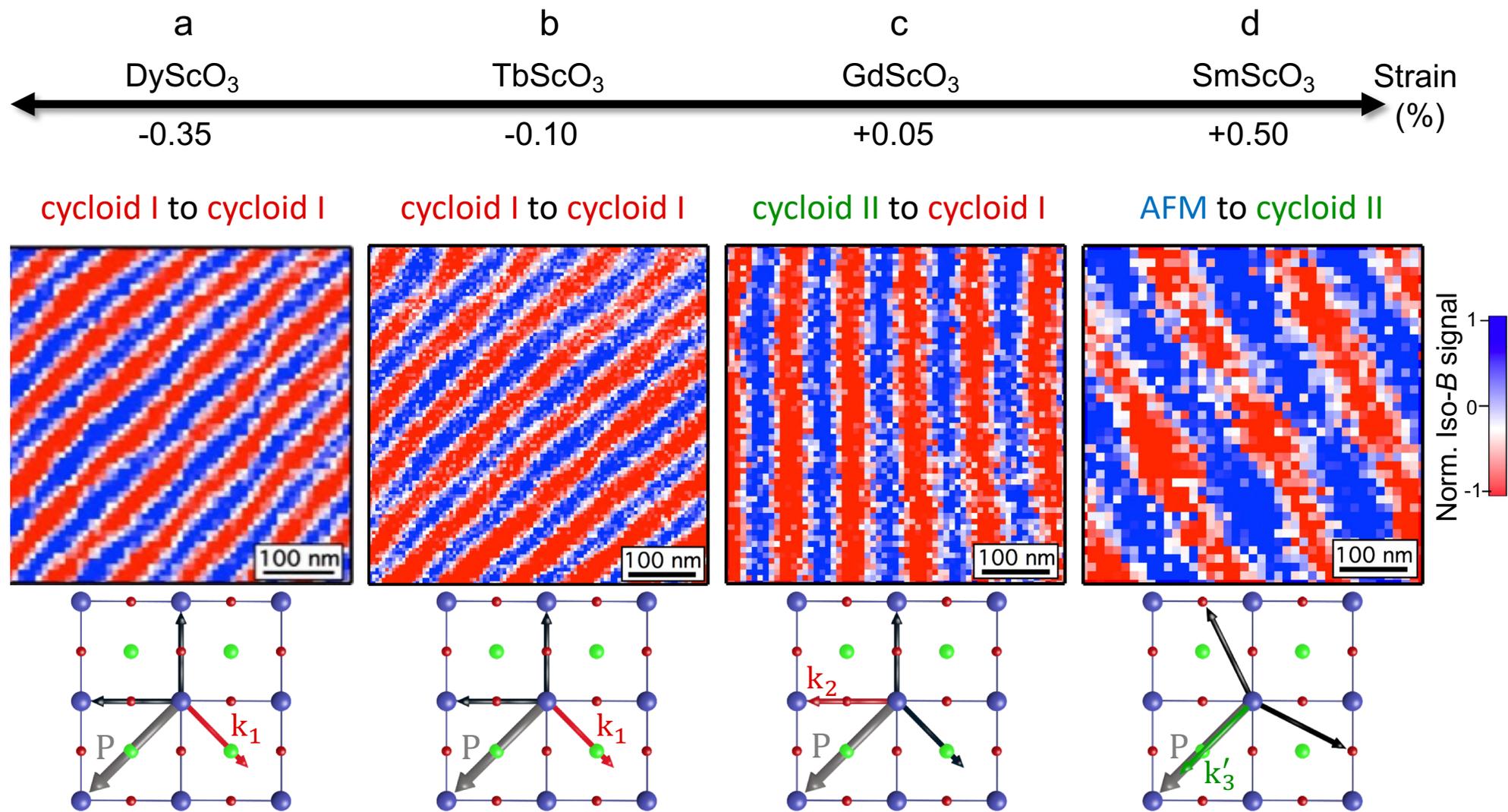

Figure 5: Magnetic textures in single ferroelectric domains as a function of strain

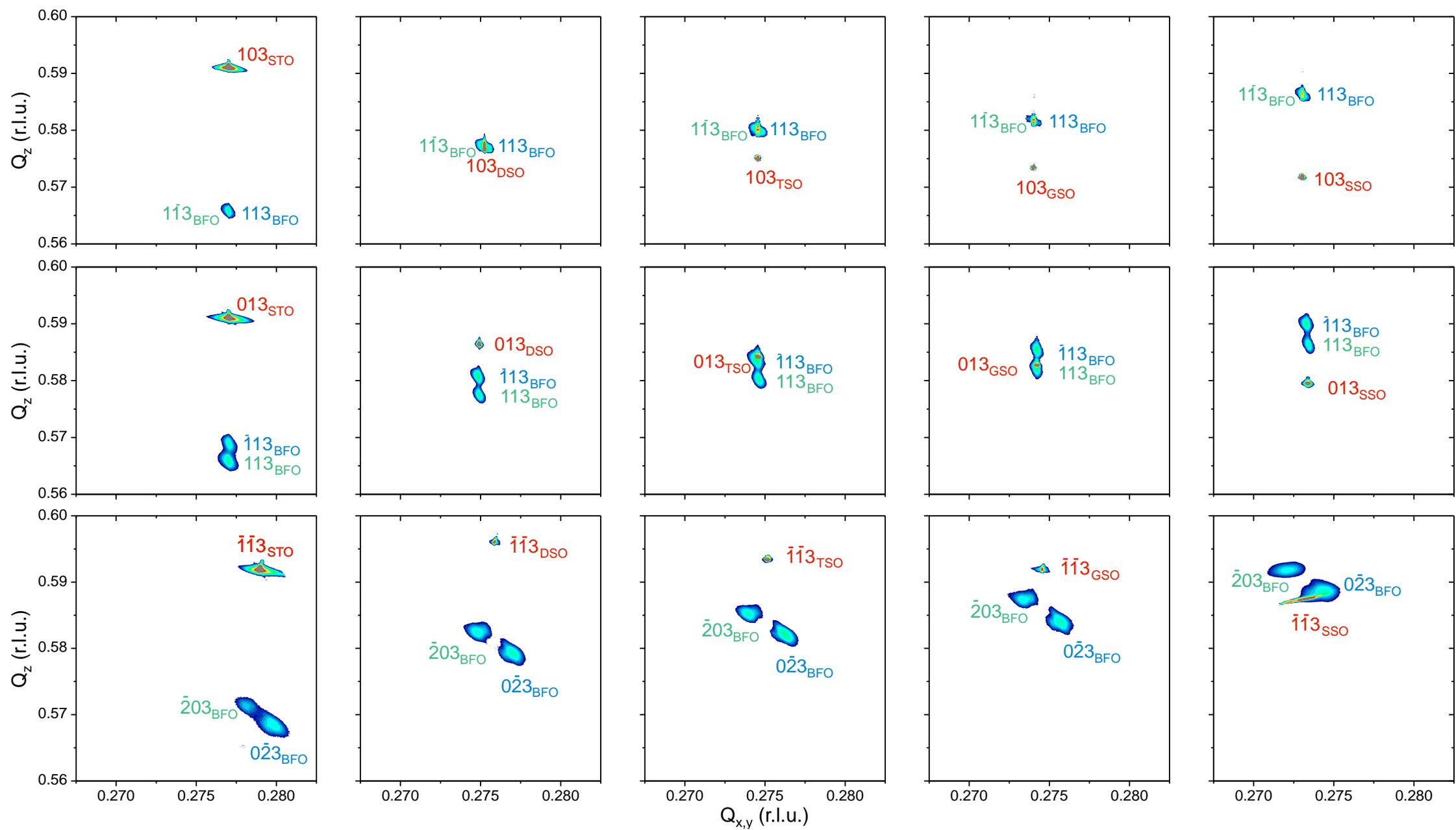

Supplementary Figure 1: RSMs on (103), (013) and (-1-13)

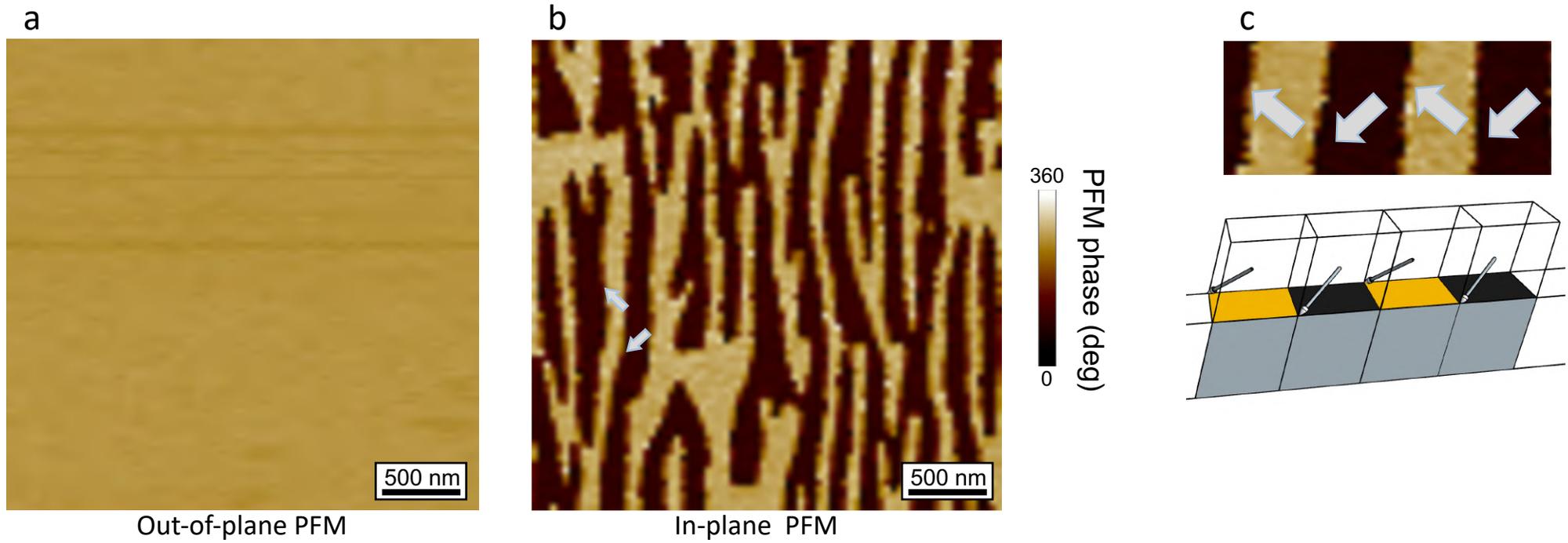

Supplementary Figure 2. Striped ferroelectric domains in BiFeO$_3$ with 71-degree domain walls.

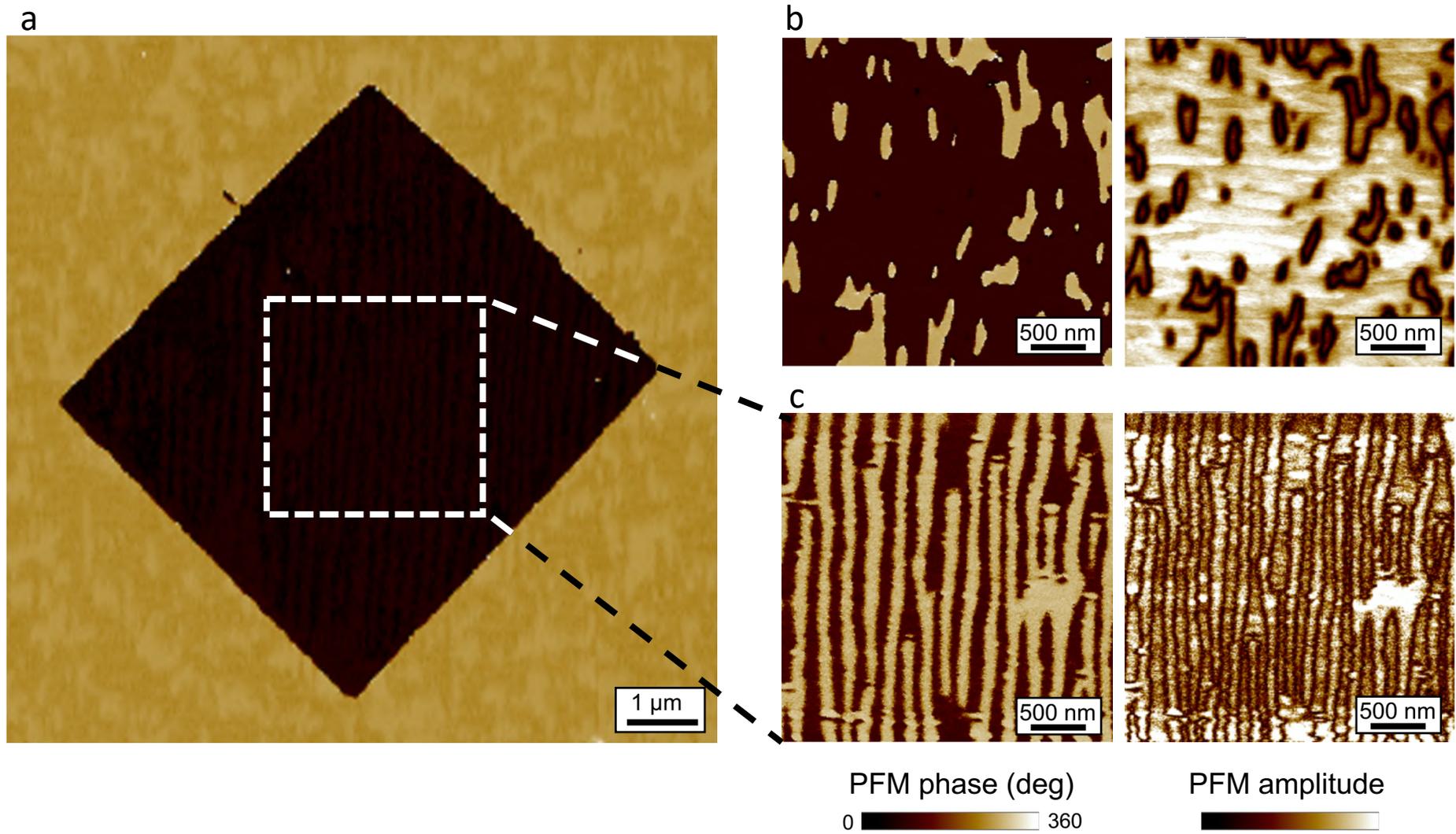

Supplementary Figure 3. Artificial stripes designed by PFM on BiFeO$_3$ thin films grown on SrTiO$_3$